\documentclass[twocolumn, numberedappendix, twocolappendix,apj, iop, a4paper]{openjournal}
\usepackage{amsmath,amsfonts,amssymb}
\usepackage[caption=false]{subfig}
\usepackage[dvipsnames,usenames]{xcolor}
\usepackage{ulem}
\usepackage{fancyhdr}
\usepackage{comment}

\definecolor{linkcolor}{rgb}{0.0,0.3,0.5}
\usepackage[
unicode, 
colorlinks=true,
linkcolor=linkcolor,
citecolor=linkcolor,
filecolor=linkcolor,
urlcolor=linkcolor,
hyperfootnotes=true,
]{hyperref}

\usepackage{epsfig}
\usepackage{graphicx}
\usepackage{bm}
\usepackage{float} 
\usepackage{color}

\begin{document}

\title{What no one has seen before: gravitational waveforms from warp drive collapse}

\author[0000-0001-8841-1522]{
 Katy Clough$^{1,2}$}
\author[0000-0003-2374-307X]{Tim Dietrich$^{3,4}$}
\author[0000-0003-4953-5754]{Sebastian Khan$^{5}$}

\address{$^1$ School of Mathematical 
Sciences, Queen Mary University of London, Mile End Road, London E1 4NS, 
United Kingdom}
\address{$^2$ Astrophysics, University of Oxford, DWB, Keble Road, Oxford OX1 3RH, UK}
\address{$^3$ Institut f\"ur Physik und Astronomie, Universit\"at Potsdam, Haus 28, 
              Karl-Liebknecht-Str. 24/25, 14476, Potsdam, Germany}
\address{$^4$Max Planck Institute for Gravitational Physics (Albert Einstein Institute), Am M\"uhlenberg 1, Potsdam 14476, Germany}
\address{$^5$ School of Physics and Astronomy, Cardiff University, Queens Buildings, Cardiff, CF24 3AA, United Kingdom}


\begin{abstract}
Despite originating in science fiction, warp drives have a concrete description in general relativity, with Alcubierre first proposing a spacetime metric that supported faster-than-light travel. Whilst there are numerous practical barriers to their implementation in real life, including a requirement for negative energy, computationally, one can simulate their evolution in time given an equation of state describing the matter. In this work, we study the signatures arising from a warp drive `containment failure', assuming a stiff equation of state for the fluid. We compute the emitted gravitational-wave signal and track the energy fluxes of the fluid. Apart from its rather speculative application to the search for extraterrestrial life in gravitational-wave detector data, this work is interesting as a study of the dynamical evolution and stability of spacetimes that violate the null energy condition. Our work highlights the importance of exploring strange new spacetimes, to (boldly) simulate what no one has seen before.
\end{abstract}

\maketitle

\section{Introduction}
Detections of gravitational waves by the LIGO Scientific-Virgo-Kagra Collaborations~\citep[e.g.,][]{LIGOScientific:2018mvr,LIGOScientific:2020ibl,KAGRA:2021vkt} are paving the way to viewing extreme gravitational events across the entire visible universe. It is natural to ask what new signals originating from strongly distorted regions of spacetime could be seen in the future beyond the compact binary mergers already detected. To search for such signals and to correctly identify them in the measured data, it is important to understand their phenomenology and properties. Numerical relativity (NR) is one of the main tools for computing gravitational-wave signals for spacetimes in the highly dynamical strongly gravity regime in the absence of symmetries \citep{Gourgoulhon:2007ue,Alcubierre:2008co,Baumgarte:2010,Baumgarte:2021skc}. NR simulations are time domain evolutions of a given initial state that allow the resulting gravitational-wave signals to be extracted in the asymptotic radiation zone, from which we can connect to our own observations from a given viewing angle and distance.

Unlike black holes, which have a natural formation channel \citep{Chandrasekhar:1931ih,Penrose:1964wq} and considerable observational support \citep{EventHorizonTelescope:2019dse,LIGOScientific:2016aoc,Ghez:2008ms}, more exotic spacetimes such as wormholes or warp drives \citep{Lobo:2007zb,Santiago:2021xjg} are considered science fiction because their formation and existence generate many potential paradoxes, and require matter that violates certain energy conditions. Nevertheless, warp drives can be described classically within the theory of general relativity by a metric first proposed by Miguel Alcubierre~\citep{Alcubierre:1994tu} and further developed in subsequent works \citep[e.g.,][]{Natario:2001tk,VanDenBroeck:1999sn,Krasnikov:1995ad,Everett:1997hb}.

The principle idea behind a warp drive is that instead of exceeding the speed of light directly in a local reference frame, which would violate Lorentz invariance, a ``warp bubble'' could traverse distances faster than the speed of light (as measured by some distant observer) by contracting spacetime in front of it and expanding spacetime behind it \citep{Alcubierre:2017kqf}. The problems associated with such spacetimes are well documented ~\citep{Everett:1995nn,Hiscock:1997ya,Visser:1998ua,Loup:2001vr,Lobo:2007zb,Obousy:2007kc}. One particular issue is the requirement for matter that violates the Null Energy Condition (NEC) \citep{Olum:1998mu,Lobo:2004wq}
\footnote{Recent works that have suggested positive energy warp drive solutions \citep{Lentz:2020euv,Bobrick:2021wog,Fell:2021wak} have problems that are discussed in \citep{Santiago:2021aup}.}. The requirement that warp drives violate the NEC may be considered a practical rather than fundamental barrier to their construction since NEC violation can be achieved by quantum effects and effective descriptions of modifications to gravity \citep{Barcelo:2002bv,Kontou:2020bta}, albeit subject to quantum inequality bounds \citep{Pfenning:1997wh,Fewster:1998xn,Krasnikov:2002dq} and other semiclassical considerations \citep{Finazzi:2009jb} that seem likely to prove problematic. Other issues with the warp drive metric include the potential for closed time-like curves \citep{Everett:1995nn,Visser:1998ua} and, from a more practical perspective, the difficulties for those in the ship in controlling and deactivating the bubble \citep{Alcubierre:2017pqm}.

From the perspective of simulating the warp drive dynamically, the key challenge is stability. One can straightforwardly generate the initial metric that describes the Alcubierre metric and infer the instantaneous matter configuration that must support it using the Einstein Equation, but a time-domain evolution of the coupled matter-gravity system necessitates that we specify the equation of state of the matter, which in turn determines how it evolves. In a perfect fluid, this involves specifying the relation between the fluid pressure and density, but in a more general case, the full relation between the stress-energy components must be given. There is (to our knowledge) no known equation of state that would maintain the warp drive metric in a stable configuration over time - therefore, whilst one can require that initially, the warp bubble is constant, it will quickly evolve away from that state and, in most cases, the warp fluid and spacetime deformations will disperse or collapse into a central point.

This instability, whilst undesirable for the warp ship's occupants, gives rise to the possibility of generating gravitational waves. The Alcubierre metric for a constant velocity warp ship would not generate gravitational waves. In addition, far away from the warp bubble, the spacetime is exactly Minkowskian -- that is, it is actually flat, not just asymptotically flat. The warp bubble spacetime is therefore undetectable by its gravitational effects as it has no gravitational wave content and no measurable ADM mass \citep{Schuster:2022ati}
\footnote{It circumvents theorems about all asymptotically Minkowskian spacetimes being Minkowski \citep{Witten:1981mf,Schon:1979rg} by the presence of the negative energy contribution. Note that the spacetime at spatial infinity, where the ADM mass is defined, is causally disconnected from the spacetime of the warp bubble, therefore the true ADM mass should be conserved and remain at zero. When we refer to the mass of the spacetime volume here, after a period of dynamical evolution, we mean a quasi-local analogy obtained by integrating over the surface of a finite volume some distance away from the collapsing bubble, after that volume has reached a stationary state with no net fluxes in or out.} (although other signatures are possible, see e.g. \citep{Marzlin:2022hfb}).
However, the collapse of the warp drive, or its acceleration or deceleration, should generate gravitational waves, and in this work, we study the first case by allowing the warp bubble to collapse. Physically, this could be related to a breakdown in the containment field that the post-warp civilisation (presumably) uses to support the warp bubble against collapse.

The prospect of a zero ADM mass spacetime radiating gravitational-wave energy immediately leads to an interesting physical question: Does the spacetime actually radiate energy, and if so, once it settles down to a steady state, does the final spacetime have a negative mass? In this work we consider the fluxes of energy both from the matter and the gravitational waves out of a spatial volume after the warp drive collapse. This study has connections to early work by Bondi \citep{Bondi:1957zz}, in which positive and negative matter restricted to linear motion was shown to attract and repel in a way that would accelerate them together.

For our study, we restrict ourselves to a limited and simplified scenario and sub-light speeds as a proof of principle. This is also a first step towards studying what, if any, pathologies develop as we reach $v=c$. We would also emphasise that the warp drive is interesting since it uses gravitational distortion of spacetime to propel the ship forward, rather than a usual kind of fuel/reaction system. This means that its detection would signal the use of NEC violating drives, even in sub-light-speed cases.

The initial bubble is described by the original Alcubierre metric with a fixed wall thickness. We develop a formalism to describe the warp fluid and its evolution and vary its initial velocity at the point of collapse (which is related to the amplitude of the warp bubble). We neglect the existence of the warp ship itself \citep{Lobo:2004wq}, which is assumed to be negligible compared to the bubble.

This article is structured as follows. In Sec.~\ref{sec-formalism}, we set out the formalism used to perform our simulations of warp bubble collapse, particularly the treatment of the warp fluid element and its equation of state. 
In Sec.~\ref{sec-results}, we present the resulting gravitational-wave signatures and quantify the radiation of energy from the spacetime region. 
In Sec.~\ref{sec-discuss}, we discuss the implications of our findings and propose directions for further study.
Some technical details and validation of our simulations are given in the Appendix.

\begin{figure}[t]
\centering
\includegraphics[width=0.95\columnwidth]{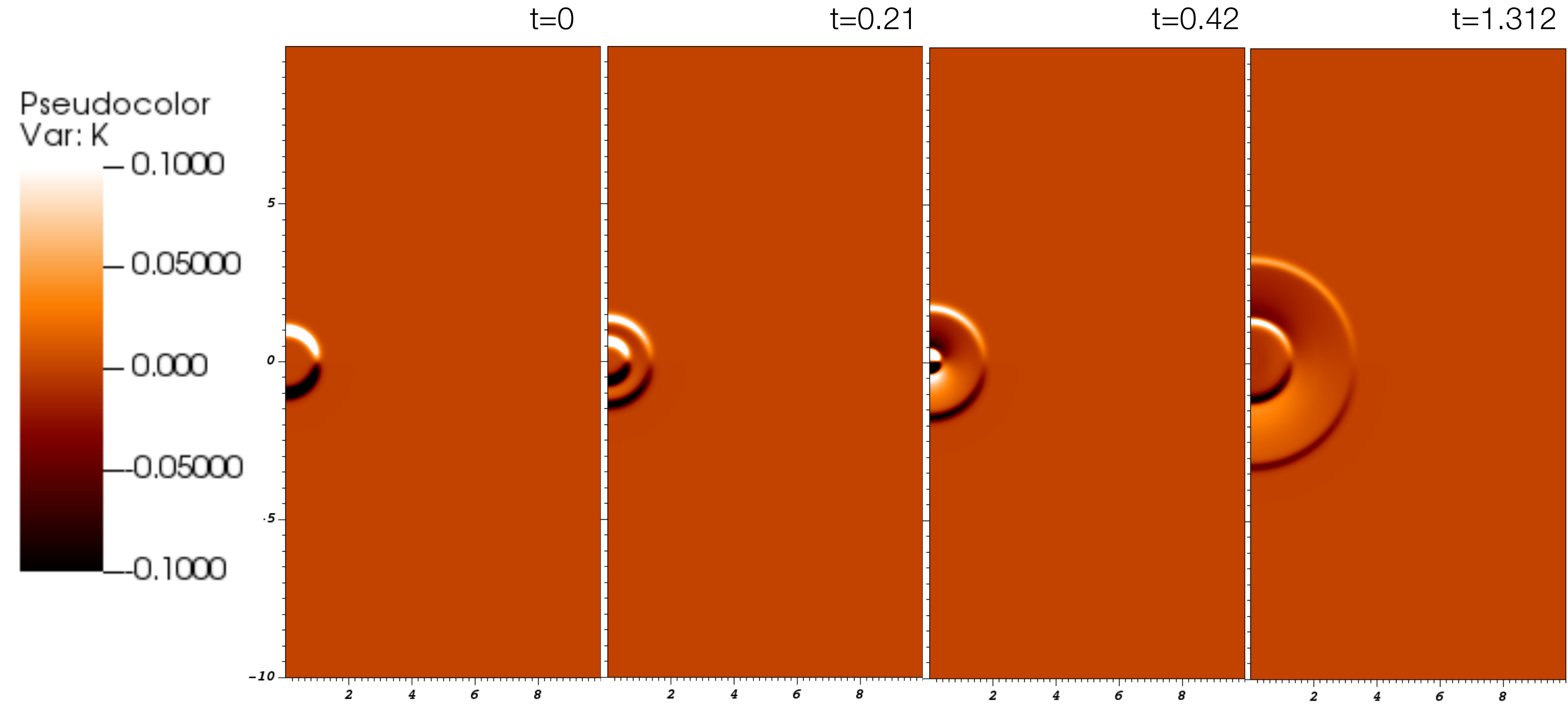}
\caption{The evolution in time of the trace of $K_{ij}$, $K=\gamma^{ij} K_{ij}$. This quantity initially provides a schematic representation of the warp drive metric, with positive values for the contracting part of the space ahead of the ship and negative values for the expanding part behind. We observe the destabilisation of these regions, which initially collapse inwards and then dissipate to infinity. Movie at \url{https://youtu.be/rZsre-y4hlM?feature=shared}.}
\label{fig:Ktrace}
\end{figure}

\section{Formalism for the warp drive simulation}
\label{sec-formalism}

We follow the standard numerical-relativity approach to solving the Einstein Equation
\begin{equation}
  G_{\mu \nu}= R_{\mu \nu} - \frac{1}{2} R g_{\mu \nu} = 8 \pi T_ {\mu \nu} 
  \label{eq:fieldeq}
\end{equation} 
as an initial value problem, using the $3+1$-decomposed form and setting $G=c=1$. This means that we define spacelike hypersurfaces that foliate the four-dimensional manifold, specify initial data on a particular hypersurface, and evolve it forward in our chosen time coordinate using the method of lines. 
(See the standard numerical relativity texts ~\citep{Arnotwitt:1960,York:1979,Alcubierre:2008co} for detailed discussions.)
The $3+1$-decomposed metric in coordinates adapted to the slicing takes the form
\begin{equation}
  {\rm d} s^2 = (-\alpha^2 + \beta_i \beta^i) {\rm d} t^2 + 2 \beta_i {\rm d}t {\rm d}x^i + 
  \gamma_{ij} {\rm d}x^i {\rm d}x^j,
\end{equation} 
where~$\alpha$ is the lapse function, $\beta^i$ is the shift vector, 
and~$\gamma_{ij}$ is the spatial metric connected to the spatial hypersurface. The description is completed by the extrinsic curvature $K_{ij}$, which is roughly related to the time derivative of $\gamma_{ij}$, and more rigorously defined as its Lie derivative, such that
\begin{equation}
K_{ij} = - \frac{1}{2 \alpha} \partial_t \gamma_{ij} + D_{(i} \beta_{j)}.
\label{eq:ext_curv}
\end{equation}
It will also be useful to note that $n_\mu = (-\alpha, 0,0,0)$ is the normal to the spatial hypersurface.

\subsection{Initial data for the metric}

Although, as discussed above, many alternatives with improved physical properties have been discussed in the literature, we will consider in this work the original Alcubierre metric~\citep{Alcubierre:1994tu} for which
\begin{eqnarray}
 \alpha      & = & 1, \\
 \gamma_{ij} & = & \delta_{ij}, \\
 \beta^i     & = & \left(0, 0, -v_s f(r_s)\right), 
\end{eqnarray}
with the quantities $r_s$ and $v_s$ defined as
\begin{equation}
 v_s  =  \frac{{\rm d}z_s}{{\rm d}t}, \quad
 r_s  = \sqrt{x^2+y^2+(z-z_s)^2},
\end{equation}
and 
\begin{equation}
 f(r_s) = \frac{\tanh(\sigma (r_s +R) )-\tanh(\sigma (r_s -R) )}{2 \tanh(\sigma R)}.
\end{equation}
The free parameters $R$ and $\sigma$ determine the shape of the warp bubble, specifically its radius and the wall thickness $\sim 1/\sigma$. 

We will use the value of $R$ to define the system of units in which our results are expressed - our simulations have a scaling freedom that means that they apply to any similar warp drive with size $R$. Specifying the physical value of $R$ determines the other physical values of our measurements, including the gravitational-wave fluxes and the energy radiated. For reference, for a warp bubble of size $R=1$ km, the relevant quantities in physical units are $t=1[R]$ is 3.33 $\mu$s and $E$ or $M=1[R]$ is 0.677 $M_\odot$
\footnote{For comparison, the Enterprise-E of the Star Trek universe is 685.7 meters in length \citep{drexler2014ships}.}.

The extrinsic curvature carries the information about the expansion and contraction of the spacetime, with values
\begin{equation}
 K_{ij} = \frac{1}{2} \left( \partial_i \beta_j + \partial_j \beta_i \right)
\end{equation}
that can be computed analytically from the expression for $f(r_s)$. The trace of $K_{ij}$, $K=\gamma^{ij} K_{ij}$ is initially $K=\partial_z \beta^z$. This quantity is usually plotted to provide a schematic representation of the warp drive, see Fig. \ref{fig:Ktrace}.

\subsection{Initial data for matter}

The initial data for the matter degrees of freedom is computed from the energy-momentum-tensor that supports the initial spacetime curvature deformation. 
We describe the state of the matter at any time in terms of projections of the stress-energy tensor
\begin{eqnarray}
S_{\mu \nu} & =& P^\sigma_\mu P^\rho_\nu T_{\sigma \rho}, \\
S_\mu& = & - P^\sigma_\mu n^\rho T_{\sigma \rho}, \\
\rho & = & n^\mu n^\nu T_{\mu \nu}, 
\end{eqnarray}
with the projection operator $P^\mu_\nu = \delta^\mu_\nu + n^\mu n_\nu$ where $n^\mu$ is the normal vector to the spatial hypersurface defined above. In coordinates adapted to the slicing, we can work with the ten components
\begin{eqnarray}
S_{ij} & =& T_{ij}, \\
S_i& = & -\frac{1}{\alpha} T_{it} + \frac{\beta^j}{\alpha}T_{ij}, \\
\rho & = & \frac{1}{\alpha^2} T_{tt} - 2 \frac{\beta^i}{\alpha^2}T_{it} + \frac{\beta^i\beta^j}{\alpha^2}T_{ij}.
\end{eqnarray}
Knowing that the energy-momentum-tensor has to satisfy the Einstein Equation, we can directly calculate the quantities on the right-hand side from Eq.~\eqref{eq:fieldeq} as
\begin{equation}
 T_{\mu \nu} = \frac{1}{8 \pi} \left( R_{\mu \nu} - \frac{R}{2} g_{\mu \nu} \right),
\end{equation}
where the Ricci tensor can be straightforwardly calculated from the Alcubierre metric in terms of derivatives of the shift. 

\subsection{Metric evolution}

\begin{figure}[t]
\centering
\includegraphics[width=0.95\columnwidth]{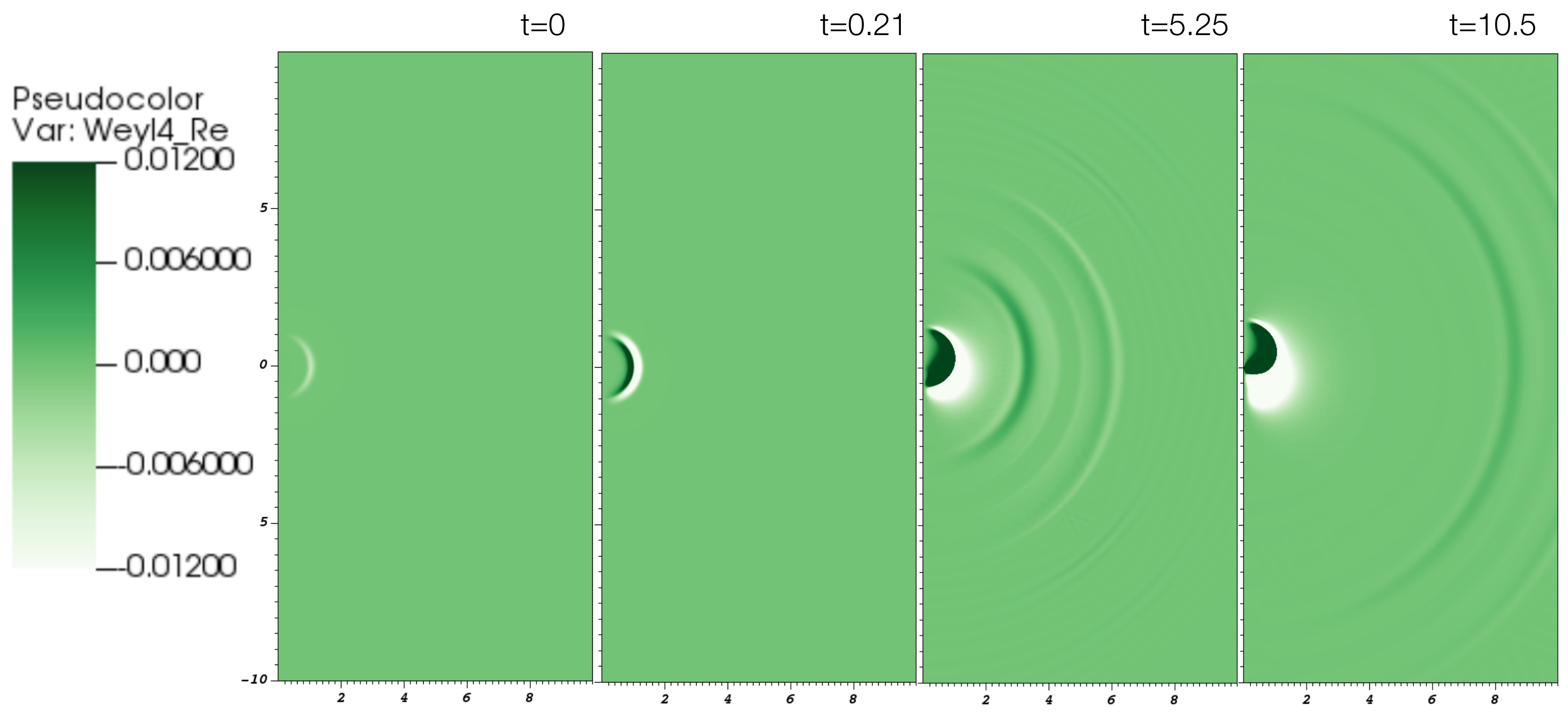}
\caption{The evolution in time of the real part of the Weyl scalar $\Psi_4$ for the case of $v=0.1$. This quantity provides a measure of the spacetime curvature and, in the far region, can be identified with the gravitational wave content of the spacetime. In the last two panels, we see a burst of gravitational-wave radiation leaving the collapsed remnant of the warp bubble. Movie at \url{https://youtu.be/rZsre-y4hlM?feature=shared}.}
\label{fig:Weyl}
\end{figure}

For the metric evolution, we use the standard CCZ4 formulation of \citep{Alic:2011gg} and the moving puncture gauge \citep{Bona:1994dr,Baker:2005vv,Campanelli:2005dd,vanMeter:2006vi} within the numerical-relativity code \textsc{grchombo} \citep{Andrade:2021rbd,Radia:2021smk,Clough:2015sqa}, which uses the method of lines, with an RK4 time integration and 4th order finite difference stencils for calculating spatial gradients.

We note that for the gauge evolution we use the shock-avoiding Bona-Masso type gauge slicing originally proposed by Alcubierre \citep{Alcubierre:1996su,Alcubierre:2002iq}, and recently advocated for by Baumgarte and Hilditch \citep{Baumgarte:2022ecu} in the context of critical collapse simulations. We find that the shock-avoiding slicing gives much better stability and, more importantly, allows us to extract coherent gravitational waveforms in the wave zone, whereas in the standard gauge, we obtained large unphysical gauge effects that hindered us in extracting the physical gravitational-wave signal. Because this gauge has a tendency to push the extrinsic curvature to zero, one should not over-interpret the evolution shown in Fig.~\ref{fig:Ktrace}. It quickly decays to zero in the simulation before the matter and gravitational-wave fluxes have begun to leave the volume. 

Better quantities to track the evolution of the metric curvature are the Weyl scalars, in particular $\Psi_4$. We study the tensorial gravitational-wave modes emitted by the warp drive by extracting the Newman-Penrose scalar $\Psi_4$ with the tetrads proposed by \citep{Baker:2001sf}, projected into spin-weight $-2$ spherical harmonics,  $\psi_{lm}=\oint_{S^2}\Psi_4|_{r=r_\mathrm{ex}}\left[{}_{-2}\bar{Y}^{lm}\right]\,\rmd \Omega$, 
where $\rmd \Omega = \sin\theta\,\rmd\theta\,\rmd\varphi$ is the area element on the $S^2$ unit sphere. The evolution of the real part of $\Psi_4$ is illustrated in Fig. \ref{fig:Weyl}.

\subsection{Matter evolution} 

\begin{figure}[t]
\centering
\includegraphics[width=0.95\columnwidth]{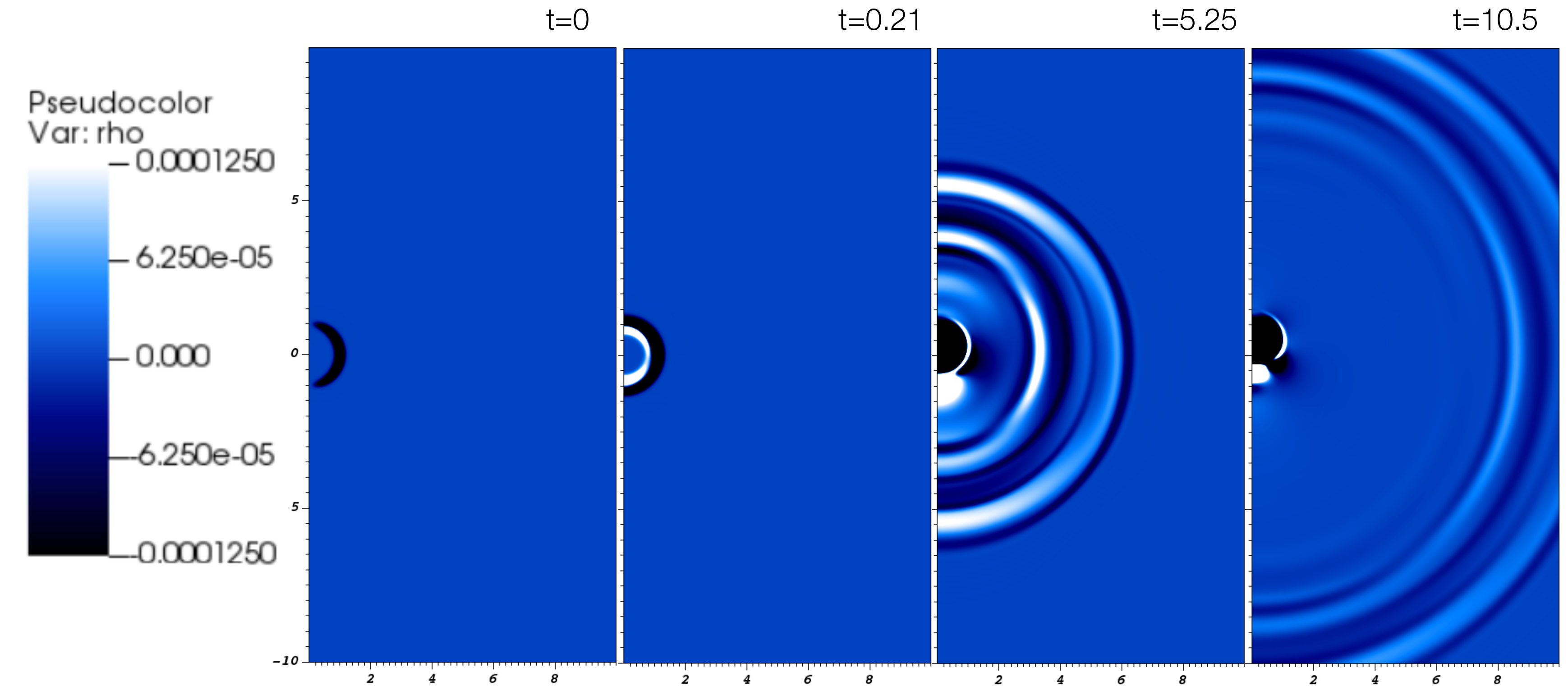}
\caption{The evolution in time of the matter energy density $\rho$ for the case of $v=0.1$. We see that a ring of positive energy forms within the initially negative energy density bubble and propels it outwards, which is reminiscent of \citep{Bondi:1957zz}. There are several waves of alternating positive and negative energies. In the final panel, we see that the remnant appears stable, and compared to Fig.~\ref{fig:Weyl}, we see that the matter waves propagate at roughly the same speed as the gravitational waves. Movie at \url{https://youtu.be/rZsre-y4hlM?feature=shared}.}
\label{fig:Rho}
\end{figure}

The requirement that $\nabla_\nu T^{\mu\nu} = 0$ gives us four evolution equations for the components of the matter $\rho$ and $S_i$ as follows: 
\begin{equation}
 \partial_t \rho =  \beta^i D_i \rho - \gamma^{ij} \frac{1}{\alpha} D_j \left( \alpha^2 S_i \right) + \alpha ( S_{ij} K^{ij} + \rho K),
\end{equation}
\begin{equation}
 \partial_t S_i =  \beta^j D_j S_i + S_j D_i \beta^j - \gamma^{jk}D_k (\alpha S_{ij}) + \alpha K S_i - \rho \partial_i \alpha.
\end{equation}

The evolution equation for $S_{ij}$ is, in principle, free - once we specify the equation of state of the matter, it would be fully determined. We base our choice on the observation that the initial warp drive spacetime has an average pressure of \citep{Santiago:2021aup}
\begin{equation}
\bar{p} \equiv \gamma^{ij} S_{ij} /3 = \rho - \frac{1}{12\pi} \nabla_a(K n^a) ~, \label{eq:pbar_warp}
\end{equation} 
which we can reexpress using the ADM formalism as
\begin{equation}
\bar{p} = \frac{\rho}{3} - \frac{1}{24\pi} \left( \frac{D^2\alpha}{\alpha} + K_{ij} K^{ij} + K^2 \right) ~.
\end{equation} 
We then decompose the stress density tensor as
\begin{equation}
    S_{ij} = \bar{p} \gamma_{ij} + \sigma_{ij}
    \label{eq:pbar}
\end{equation}
where the second term is trace-free by construction and represents the anisotropic stress component (it is initially non-zero for the Alcubierre metric). We then impose that
\begin{eqnarray}
\partial_t \sigma_{ij} &=&  
    \beta^k D_k \sigma_{ij} + \sigma_{kj} \partial_i \beta^k + \sigma_{ik} \partial_j \beta^k - \kappa \alpha \sigma_{ij}, \\
    \partial_t \bar{p} &=&  
    \beta^i \partial_i \bar{p}
    + \partial_t {\rho}
    - \beta^i \partial_i \rho
    - \kappa \alpha (\bar{p} - \rho),
\end{eqnarray}
where the first terms are the advection of the fields in spacetime and the terms containing $\kappa$ set a timescale related to proper time for the decay of the anisotropic stresses to zero and the pressure to that of a stiff fluid for which $\bar{p} = \rho$. We take the timescale $1/\kappa$ to be of order $[R]$. Evolving the anisotropic and average pressure components, we can at any point reconstruct $S_{ij}$ from their values as defined by Eq.~\eqref{eq:pbar}\footnote{One might consider trying to naively evolve the equation of state in Eq.~\eqref{eq:pbar_warp}, but a simple analysis shows that this introduces higher derivative terms to the equation of motion for the momentum density (either 3rd spatial derivatives of the lapse or 3rd-time derivatives of the spatial metric). Given the high chance that such terms will result in an ill-posed evolution system, we do not follow this approach.}.

This construction assumes that there is some transition after the warp field breakdown that reverts the warp fluid to a perfect fluid with a stiff equation of state and a velocity that matches the normal observers, such that the divergence term in Eq.~\eqref{eq:pbar_warp} and the anisotropic stresses decay away. Further investigations are needed to understand the influence of the equation of state of the warp field on the gravitational-wave signal, but it appears that the metric behaviour mostly dominates the signal. However, the chosen equation of state will strongly affect the matter fluxes, which may change the conclusions in Sec. \ref{sec-results}.

The evolution of the energy density $\rho$ that results from this prescription is illustrated in Fig. \ref{fig:Rho}.

\subsection{Numerical setup}

The size of the domain is $L=168~[R]$ with $N=224$ cells on the coarsest grid in the direction of travel of the warp ship (the $z$ axis). In the $x$ and $y$ directions, we use the symmetry of the problem to halve the simulation domain so that we only evolve a quarter of the full domain, with appropriate parities at the reflection boundaries. In the nonsymmetric directions, we use Sommerfeld boundary conditions ~\citep{Radia:2021smk}. We employ 6 levels of 2:1 refinement, such that the finest grid has a width of $2.625 [R]$ and the finest grid spacing is $dx\sim 0.0117 [R]$. 

For all runs, the size of the warp ship in code units is set to be $R=1$, and the value that determines the wall thickness $\sigma$ is set to be $8$. The velocity of the warp ship is varied between 0.1 and 0.2. We have found that we can stably evolve higher velocities up to 0.5, but good quality results require very high resolutions, which would be better achieved with a code adapted to the axisymmetry. This work can be viewed as a step towards the study of the physicality of the faster-than-light case, where there are good reasons to suppose that the spacetime should be pathological in some way. By showing that numerical evolutions can be made to work at sub-light speeds, this gives us hope of being able to trace a point of breakdown near light speed.

The waves are extracted at a radius of $14[R]$. This is much smaller than values used in binary extraction, but the system here is very different and so the comparison is not appropriate. In particular, the spacetime is already flat outside the original bubble wall at $R$. The total integrated mass of the matter is also initially of order (minus) $M = \int \rho dV \sim 0.01R$, and not $M\sim R$ as in, say, a neutron star. Since the frequency of the waves is high relative to the grid spacing, we need to extract further in to obtain a good resolution of the frequencies using the adaptive mesh refinement (AMR) hierarchy. We have checked the consistency of the waves at different resolutions, and details of the checks we have done on convergence of the constraints and verification of the matter evolution are given in Appendix \ref{app:technical}.

\section{Results}
\label{sec-results}

\subsection{Gravitational waves}

\begin{figure}[t]
\centering
\includegraphics[width=0.95\columnwidth]{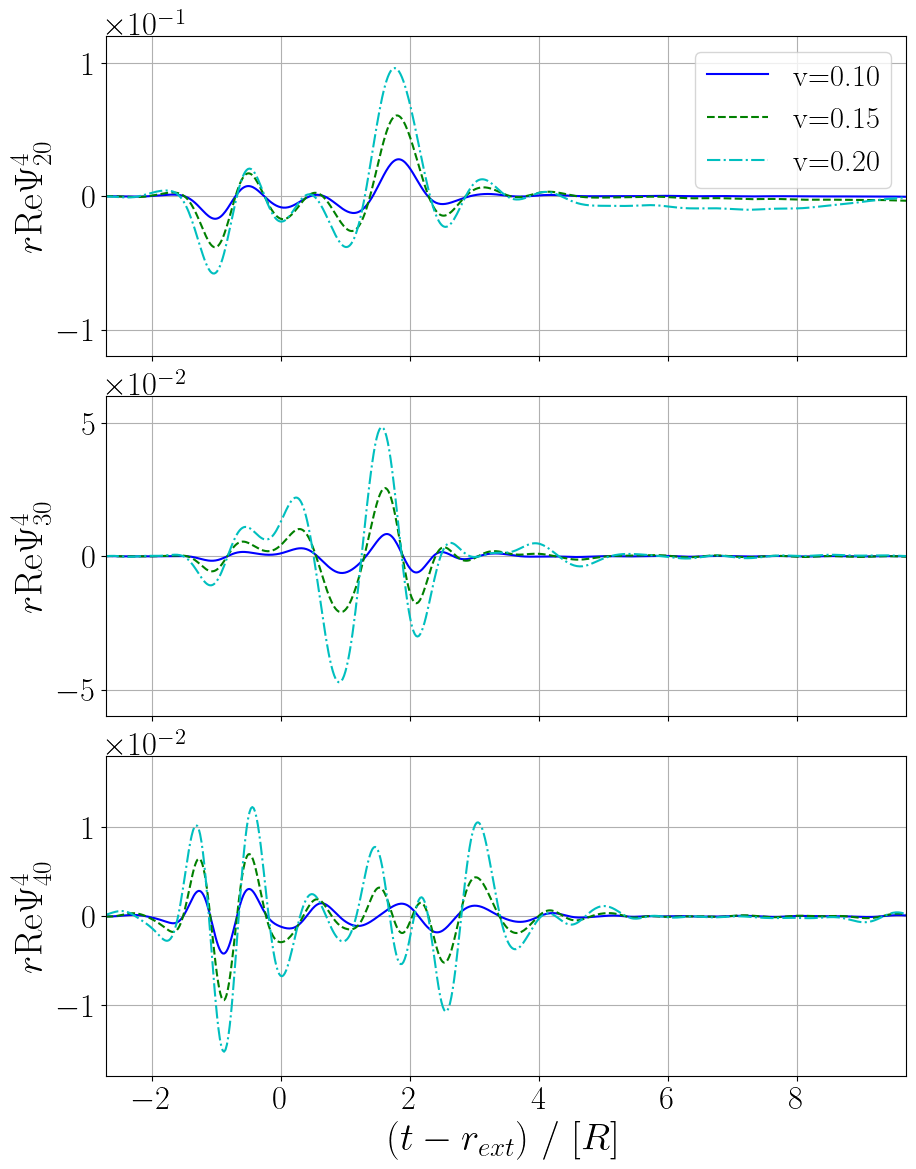}
\caption{Modes of the Newman-Penrose scalar $\Psi_4$ multiplied by the coordinate radius for different values of the warp speed $v$.  Given the symmetry of the problem, the m$\neq 0$ modes are zero, so we only show the modes with $l=2,3,4$. We observe that the signal has a frequency set by the size of the warp bubble $f \sim 1/[R]$, which for a 1 km warp bubble would be of the order $f \sim 300$ kHz.}
\label{fig:GW_Weyl}
\end{figure}

A visualisation of the evolution of the gravitational-wave content is shown in Fig.~\ref{fig:Weyl}. We plot the extracted Weyl scalar $\Psi_4$ for the main $m=0$ modes in Fig. \ref{fig:GW_Weyl}. The signal comes as a burst, initially having no gravitational wave content, followed by an oscillatory period with a characteristic frequency of order $1/[R]$. Overall, the signal is very distinct from the typical compact binary coalescences observed by gravitational-wave detectors and more similar to events like the collapse of an unstable neutron star or the head-on collision of two black holes~\citep{Giacomazzo:2011cv,Sperhake:2005uf}. However, since the remnant is not a black hole, there is no characteristic black hole ringdown in the signal, and there is evidence of a longer lower frequency tail, particularly for higher velocities.

Integrating $\Psi_4$, we find that for a warp speed of $v=0.1$, the magnitude of the strain times the radius of extraction is of order $10^{-2} [R]$. Thus, for a 1 km-sized warp ship, the strain of the signal at a distance of 1 Mpc would be $10^{-21}$. However, its frequency ($f\sim 1/[R]$) equates to $f\sim$ 300 kHz, which is outside the range of current ground based detectors. Proposals for higher frequency detectors have been made \citep{Aggarwal:2020olq}, so in future one may be able to put bounds on the existence of such signals. Depending on the size of the warp bubble, which determines the frequency of the signal, as well as the distance to the source, these signatures may be detected through a coherent search or burst search combining information from multiple gravitational-wave detectors \citep[e.g.,][]{Sutton:2009gi,Cornish:2014kda,Klimenko:2015ypf}. A more in depth analysis, like the one of \cite{Kuwahara:2023tns}, would be needed to give accurate bounds.

We note that since the matter and gravitational waves propagate at the same speed, the gravitational waves are never in an exact vacuum. We use the appropriate calculation of the Weyl scalars to account for this, but it means that the two components may continue to interact and that the gravitational waves will not fall off exactly as $1/r$ (although the difference is small over the domain we study). This is specific to the case where the matter propagates at the speed of light, whereas in a more realistic matter model we would expect the matter to be moving slower and the gravitational waves to propagate ahead. This is another point that could be investigated further in future work.

\subsection{Matter fluxes}

We measure the matter flux out of a spherical coordinate surface $\partial\Sigma$ with unit coordinate normal vector $s^i$ and coordinate area element $dA$ as
\begin{equation}
    F = \int_{\partial\Sigma} \sqrt{\gamma} ~T^i_0 s_i dA
\end{equation}
where $\gamma$ is the determinant of the spatial metric $\gamma_{ij}$ (see \cite{Clough:2021qlv} for a derivation and details on practical implementation).
We can also integrate this to find the total energy flux that is emitted up to a time $t$ as
\begin{equation}
    \Delta E = \int_0^t F(\tilde{t}) d\tilde{t} ~.
\end{equation}

\begin{figure}[t]
\centering
\includegraphics[width=0.98\columnwidth]{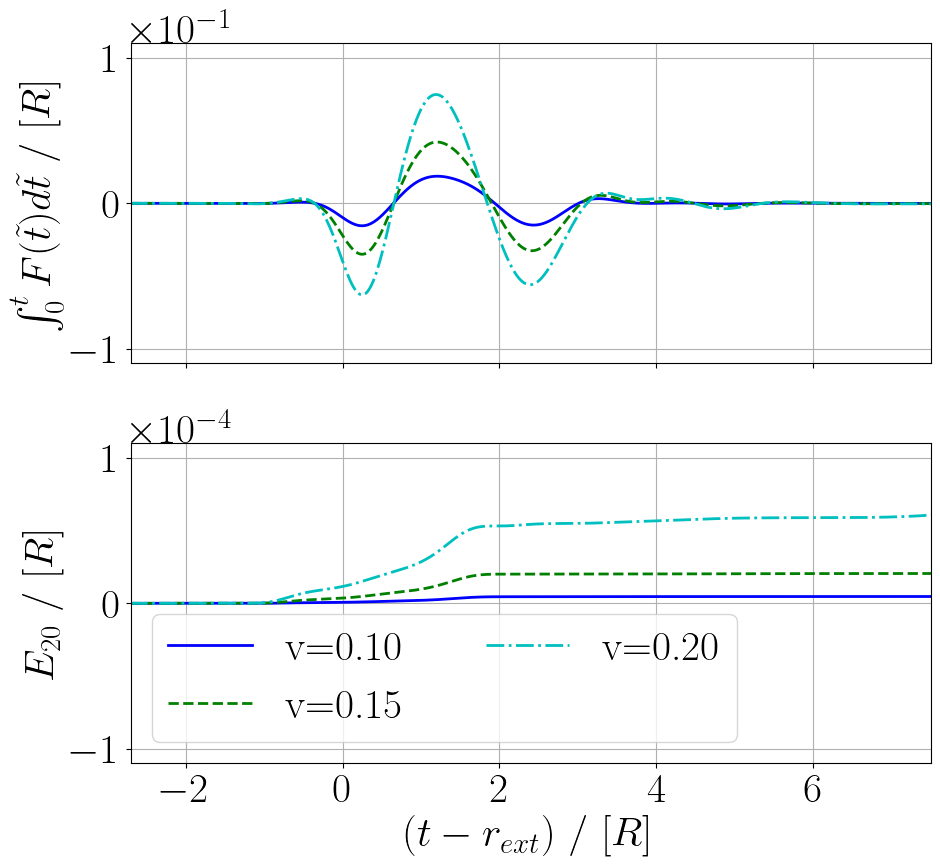}
\caption{A comparison of the integrated matter and gravitational-wave energy fluxes out of sphere at a coordinate radius of $r=14[R]$ around the warp bubble versus time for different $v$. The top panel shows the integrated matter fluxes (i.e., the total energy radiated). The bottom panel shows the integrated gravitational wave fluxes in the 20 mode. The net gravitational-wave flux is always positive, whereas the net matter flux oscillates as waves of positive and negative energy leave the volume. Note that the matter flux is significantly larger in magnitude than that of the gravitational waves.}
\label{fig:Energy}
\end{figure}

A visualisation of the evolution of the matter-energy density $\rho$ is shown in Fig.~\ref{fig:Rho}. In Fig.~\ref{fig:Energy}, we show a comparison of the integrated matter and gravitational-wave fluxes out of the sphere at a coordinate radius of $r=14[R]$ around the warp bubble versus time. We see that whilst the gravitational-wave flux is strictly positive as expected, the matter flux alternates in sign, with waves of positive and negative energy leaving the volume. It appears from the evolution that the ejection of matter is driven by a shell of positive energy forming within the negative energy density shell of the initial bubble and driving it outwards. This is reminiscent of the early works of \cite{Bondi:1957zz}, showing the effect of repulsion and attraction between positive and negative energy matter.

\begin{figure}[t]
\centering
\includegraphics[width=0.98\columnwidth]{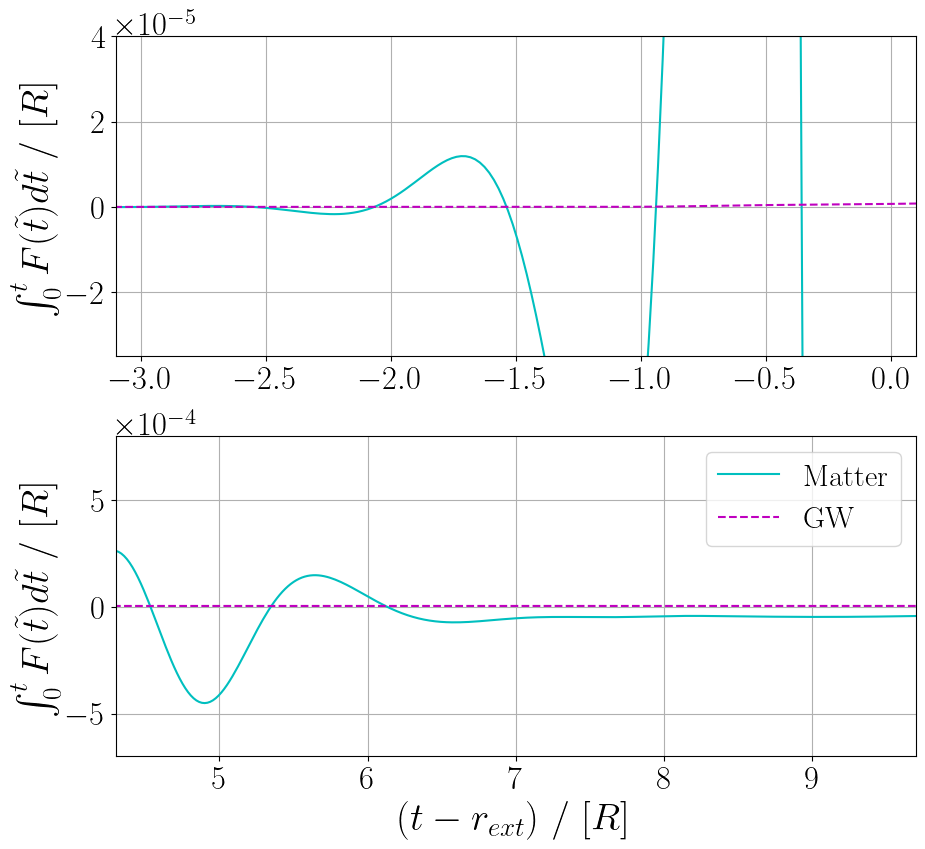}
\caption{Zoom in on the start and end periods of the integrated energy fluxes in the matter and gravitational waves as in Fig. \ref{fig:Energy} for the case $v=0.1$. We see that the flux begins in the matter sector, with alternating waves of positive and negative energy. The gravitational-wave flux starts a little later but continues throughout the periods in which the net matter flux is both positive and negative. The second panel shows that the overall matter energy loss is negative (i.e. the energy of matter in the volume increases), whereas the gravitational-wave flux is overall positive as expected but much smaller. The final mass of the spacetime volume once the spacetime settles into a stationary state} is therefore larger than its initial value (which was zero).
\label{fig:EnergyZoom}
\end{figure}

In Fig.~\ref{fig:EnergyZoom}, we show a zoom-in of the fluxes and integrated energy loss, which shows that the matter flux results in an overall negative flux out of the spacetime, with the gravitational waves giving a much smaller positive flux. The result is that the final mass of the spacetime volume is more positive (having started at zero). It would be interesting to study whether this is a generic effect, such that spacetimes with NEC-violating matter cannot dynamically settle into a state with a mass below zero. We can see from Fig. \ref{fig:EnergyZoom} that nothing appears to impede the gravitational-wave flux when the net matter flux is instantaneously positive overall, as between times $t-r_{ext}=-1.0$ and $-0.5$ on the second panel of the plot so temporary decreases appear to be permitted.

Since we do not know the type of matter used to construct the warp ship, we do not know whether it would interact (apart from gravitationally) with normal matter as it propagates through the Universe. If it did interact, it may give rise to further signatures (i.e., a multimessenger event). For a 1 km sized warp bubble traveling at $v=0.1$, the magnitude of the energy carried by the matter waves, of order $10^{-2}$ [R] (see Fig. \ref{fig:Energy}), would be around 1/100 times the mass-energy of the sun.

The evolution does appear to settle down into a stable remnant, that could be described as a ``star'' of the warp fluid - it is a configuration of the fluid that appears to balance the gravitational forces and fluid pressures. We would need to evolve for longer and use a larger grid (to avoid boundary reflection effects) to be sure about its stability, but it appears to have settled to at least an approximately steady state in the period simulated.

\section{Discussion}
\label{sec-discuss}

In this work, we proposed a formalism for studying warp drive spacetimes dynamically and produced the first fully consistent numerical-relativity waveforms for the collapse of a warp drive bubble. 

We also studied the evolution of the fluxes of matter and gravitational-wave energy from the spacetime, finding that an initial wave of negative energy matter is emitted, followed by alternating positive and negative energy waves. The gravitational-wave flux starts shortly after and is positive throughout, as expected. The end result is a net negative flux of energy, resulting in a higher final mass of the spacetime volume. It would be interesting to explore to what extent this behaviour is generic or depends on the chosen equation of state supporting the warp bubble. To do this, one would want to consistently implement a fluid description with a general rest frame velocity and equation of state, unlike the somewhat ad hoc setup employed here.

This work could be extended in several other directions. Firstly, using a code better adapted to the axisymmetry, one could extend to higher speeds, including those that exceed $v=1.0$, to see if there is a change in behaviour as light speed is approached. One might expect to see some kind of pathology develop, e.g., the formation of elliptic regions that cause the dynamical evolution to fail. Better resolution and longer evolutions would also allow better characterization of the strain signal, particularly the role of the gravitational-wave memory content~\citep{Mitman:2024uss}. 

As discussed above, for a 1km-sized ship, the frequency of the signal $\sim 300$ kHz is much higher than the range probed by existing detectors, so current observations cannot constrain the occurrence of such events. However, the amplitude of the strain signal would be significant for any such event within our galaxy and even beyond - at a distance of 1 Mpc, slightly further than the Andromeda galaxy, the strain amplitude is similar to LIGO detector's peak sensitivity. The signal is therefore potentially within the reach of future detectors targeting higher frequencies \citep{Aggarwal:2020olq}. We caution that the waveforms obtained are likely to be highly specific to the model employed, which has several known theoretical problems, as discussed in the Introduction. Further work would be required to understand how generic the signatures are, and properly characterise their detectability.

\section{Acknowledgements}
The crew thanks the creators and actors of the Star Trek universe for their inspiration for this work. We also thank the many people who discussed the project and gave useful feedback and encouragement, especially the 2023 Peyresque workshop attendees and our NR friends at the 2023 ETK meeting in Aveiro.  Particular conceptual help was given by Jean Alexandre, Josu Aurrekoetxea, Mark Hannam, Juan Valiente Kroon, Gautam Satischandran, and Miren Radia and Josu Aurrekoetxea provided the gravitational-wave analysis scripts used. The \texttt{GRTL} collaboration is warmly acknowledged for the use of their GRChombo code (\texttt{www.grchombo.org}) and the development work that supports it.
Katy ``I cannae change the laws of physics Captain'' Clough thanks the University of Potsdam and the Albert Einstein Institute in Potsdam for hosting her for the final stage of the project and acknowledges current support from an STFC Ernest Rutherford fellowship, project reference ST/V003240/1, and previous support from the European Research Council (ERC) under the European Union’s Horizon 2020 research and innovation programme (grant agreement No 693024). Tim ``Captain Jim T." Dietrich thanks Pedro Ferreira for hosting him at the University of Oxford, which allowed this project to make significant progress. TD also acknowledged support from the European Union (ERC, SMArt, 101076369). Views and opinions expressed are those of the authors only and do not necessarily reflect those of the European Union or the European Research Council. Neither the European Union nor the granting authority can be held responsible for them.
Sebastian ``Wrath of" Khan thanks Jens Niemeyer for hosting him in Georg August University of Goettingen, where the voyage towards this article first began. 

The simulations presented in this paper used the Glamdring cluster in Astrophysics at Oxford University and DiRAC resources under the projects of dp092 (Testing the limits of General Relativity). This work used the DiRAC Memory Intensive service (Cosma8 and Cosma7) at Durham University, managed by the Institute for Computational Cosmology on behalf of the STFC DiRAC HPC Facility (www.dirac.ac.uk). The DiRAC service at Durham was funded by BEIS, UKRI and STFC capital funding, Durham University and STFC operations grants. This work also used the DiRAC Data Intensive service (DIaL3) at the University of Leicester, managed by the University of Leicester Research Computing Service on behalf of the STFC DiRAC HPC Facility (www.dirac.ac.uk). The DiRAC service at Leicester was funded by BEIS, UKRI and STFC capital funding and STFC operations grants. This work also used the DiRAC Data Intensive service (CSD3) at the University of Cambridge, managed by the University of Cambridge University Information Services on behalf of the STFC DiRAC HPC Facility (www.dirac.ac.uk). The DiRAC component of CSD3 at Cambridge was funded by BEIS, UKRI and STFC capital funding and STFC operations grants. DiRAC is part of the UKRI Digital Research Infrastructure.

\begin{figure}[b]
\centering
\includegraphics[width=0.95\columnwidth]{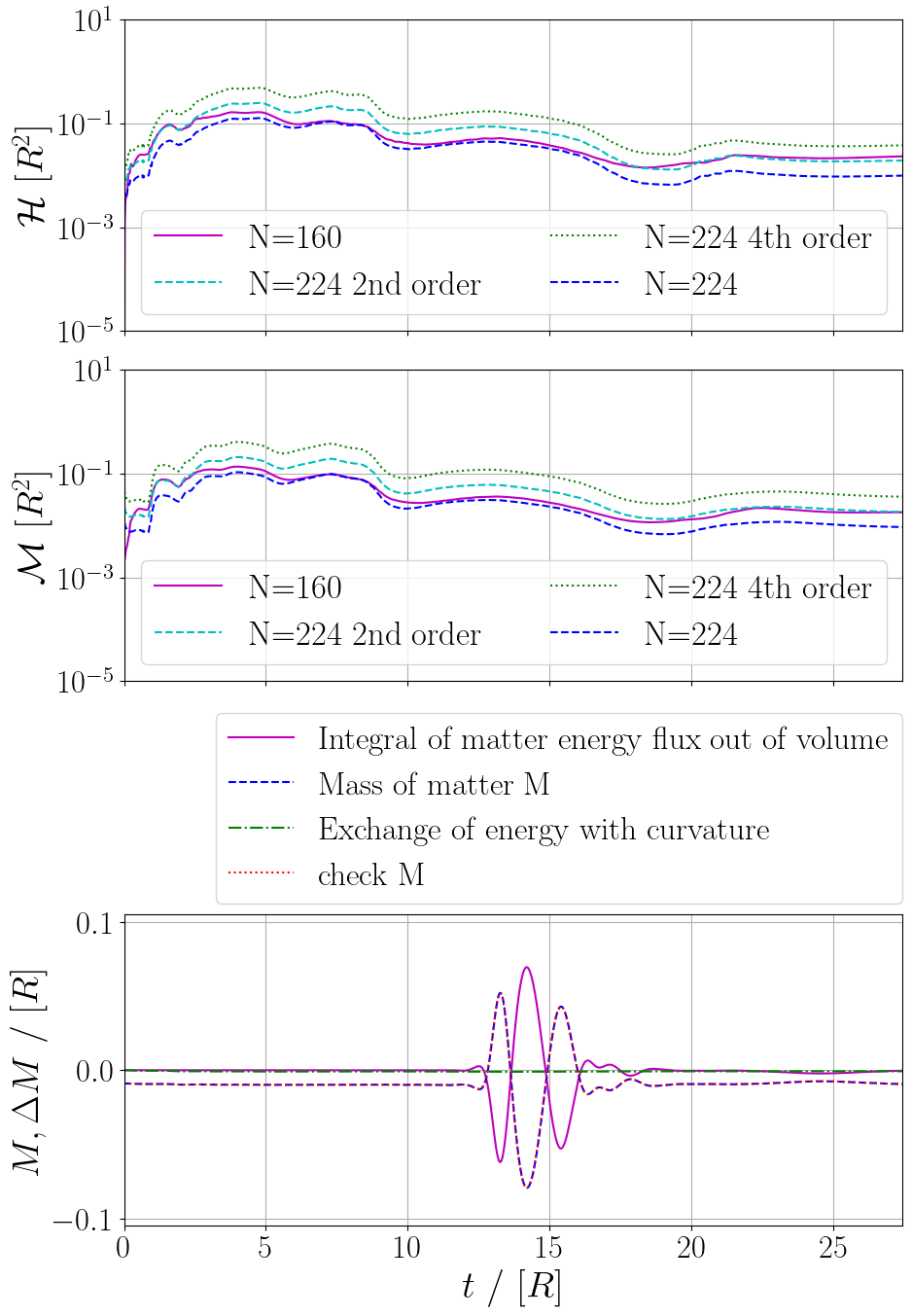}
\caption{The top panel shows the L2 norm of the Hamiltonian constraint violation over time in the central volume over which data is extracted, the second the same for the Momentum constraint violation. These show approximate convergence of the quantities to zero with increasing resolution between 1st and and 3rd order. The final panel shows the reconciliation of the matter-energy fluxes and source term (the exchange of energy with the curvature) as described in the main text. The agreement between the direct measurement of M and its inferred value from the flux and source terms validates that the matter is well resolved in the relevant volume.}
\label{fig:converge}
\end{figure}

\appendix 
\section{Numerical details, validation and convergence}
\label{app:technical}

To validate our simulations, we monitor the violations of the Hamiltonian and momentum constraints which are given by
\begin{align}
    \mathcal{H} &:= R + \frac{2}{3}K^2 
    - \tilde{A}_{kl}\tilde{A}^{kl}-2\Lambda-16\pi\rho,
    \label{eq:Ham-constraint}\\
        \mathcal{M}_i &:= \tilde{\gamma}^{kl}\left(\partial_k \tilde{A}_{li} 
        -2\tilde{\Gamma}^m_{l(i}\tilde{A}_{k)m}
        -3\tilde{A}_{ik}\frac{\partial_l\chi}{2\chi}\right) \nonumber \\
        & \quad -\frac{2}{3}\partial_iK - 8\pi S_i.
    \label{eq:mom-constraint}
\end{align}
In the continuum limit these two quantities should vanish. 
We check that the initial values converge to zero at the appropriate order, which validates the initial stress energy tensor components $\rho$ and $S_i$. The Hamiltonian constraint is solved exactly to numerical precision, whereas the spatial gradients in the Momentum constraint mean that it should initial converge at 4th order. During the evolution we expect errors to be dominated by the time integration, and give something between 1st and 3rd order, with some of the drop in convergence caused by the approximate way that we zero the constraints outside the volume.

We calculate the L2 norm of the constraints over the coordinate volume for which the matter integrals are evaluated (with its 2D surface also being the surface used for extraction of the gravitational waves), which is at a coordinate radius of $r=14[R]$ from the centre of the grid.
As shown in Fig \ref{fig:converge}, the constraint violations remain under control throughout the simulation and converge at a reasonable order.

To verify that the matter is sufficiently resolved within the volume of interest, we also reconcile the matter quantities over time, checking that the net energy flux out of the surface matches the change in the matter energy and any exchange with the curvature (often referred to as a ``source'') over time. For this, we follow the approach of \cite{Clough:2021qlv}. As shown in the last panel of Fig.~\ref{fig:converge}, the agreement is good throughout the simulation, which gives us confidence in the matter evolution.

 
\bibliographystyle{plainnat}
\bibliography{refs.bib}

\end{document}